\documentclass{jpsj2}

\newcommand{\vv}[1]{\mbox{\boldmath{$#1$}}}
\newcommand{\vs}[1]{\mbox{\footnotesize \boldmath{$#1$}}}

\def\runtitle{
Multi-Modes Phonon Softening in Two-Dimensional Electron-Lattice System} 
\def\runauthor{Shutaro \textsc{Chiba} and Yoshiyuki \textsc{Ono}}

\title{%
Multi-Modes Phonon Softening in Two-Dimensional Electron-Lattice System}

\author{%
Shutaro \textsc{Chiba}\thanks{E-mail address: shutaroc@ph.sci.toho-u.ac.jp } 
and
Yoshiyuki \textsc{Ono}\thanks{E-mail address: ono@ph.sci.toho-u.ac.jp }
}

\inst{%
Department of Physics, Toho University, Miyama 2-2-1, Funabashi, 
Chiba 274-8510 \\
}

\recdate{\today}

\abst{%
Phonon dispersion in a two-dimensional electron-lattice system described by 
a two-dimensional square-lattice version of  Su-Schrieffer-Heeger's model and 
having the half-filled electronic band is studied theoretically at 
temperatures higher than the mean field critical temperature of the Peierls 
transition. When the temperature is lowered from the higher region down to 
the critical one, softening of multi phonon modes which have wave vectors 
equal to the nesting vector $\vv{Q}=(\pi/a,\pi/a)$ with $a$ the lattice 
constant or parallel to $\vv{Q}$ is observed. Although both of 
the transverse and longitudinal modes are softened at the critical 
temperature in the case of the wave vector equal to $\vv{Q}$, only the 
transverse modes are softened for other  wave vectors parallel to $\vv{Q}$. 
This behavior is consistent with the Peierls distortions at lower 
temperatures.
}

\kword{%
two-dimensional electron-lattice system, phonon softening, Peierls transition, 
longitunidal mode, transverse mode, linear mode analysis, SSH model
}

\begin{document}
\sloppy
\maketitle

\section{Introduction} \label{sec1}

\begin{figure}[htb]
 \begin{center}
\unitlength 0.1in
\begin{picture}(22.35,20.20)(13.85,-26.20)
\put(25.9000,-16.1000){\makebox(0,0)[rt]{O}}%
\put(25.6000,-6.0000){\makebox(0,0)[rt]{$k_{y}$}}%
\put(36.8000,-16.6000){\makebox(0,0)[rt]{$k_{x}$}}%
%
\special{pn 8}%
\special{pa 2600 2620}%
\special{pa 2600 600}%
\special{fp}%
\special{sh 1}%
\special{pa 2600 600}%
\special{pa 2580 667}%
\special{pa 2600 653}%
\special{pa 2620 667}%
\special{pa 2600 600}%
\special{fp}%
%
\special{pn 8}%
\special{pa 1620 1600}%
\special{pa 3620 1600}%
\special{fp}%
\special{sh 1}%
\special{pa 3620 1600}%
\special{pa 3553 1580}%
\special{pa 3567 1600}%
\special{pa 3553 1620}%
\special{pa 3620 1600}%
\special{fp}%
\special{pn 8}%
\special{pa 1620 1600}%
\special{pa 1625 1600}%
\special{pa 1630 1600}%
\special{pa 1635 1600}%
\special{pa 1640 1600}%
\special{pa 1645 1600}%
\special{pa 1650 1600}%
\special{pa 1655 1600}%
\special{pa 1660 1600}%
\special{pa 1665 1600}%
\special{pa 1670 1600}%
\special{pa 1675 1600}%
\special{pa 1680 1600}%
\special{pa 1685 1600}%
\special{pa 1690 1600}%
\special{pa 1695 1600}%
\special{pa 1700 1600}%
\special{pa 1705 1600}%
\special{pa 1710 1600}%
\special{pa 1715 1600}%
\special{pa 1720 1600}%
\special{pa 1725 1600}%
\special{pa 1730 1600}%
\special{pa 1735 1600}%
\special{pa 1740 1600}%
\special{pa 1745 1600}%
\special{pa 1750 1600}%
\special{pa 1755 1600}%
\special{pa 1760 1600}%
\special{pa 1765 1600}%
\special{pa 1770 1600}%
\special{pa 1775 1600}%
\special{pa 1780 1600}%
\special{pa 1785 1600}%
\special{pa 1790 1600}%
\special{pa 1795 1600}%
\special{pa 1800 1600}%
\special{pa 1805 1600}%
\special{pa 1810 1600}%
\special{pa 1815 1600}%
\special{pa 1820 1600}%
\special{pa 1825 1600}%
\special{pa 1830 1600}%
\special{pa 1835 1600}%
\special{pa 1840 1600}%
\special{pa 1845 1600}%
\special{pa 1850 1600}%
\special{pa 1855 1600}%
\special{pa 1860 1600}%
\special{pa 1865 1600}%
\special{pa 1870 1600}%
\special{pa 1875 1600}%
\special{pa 1880 1600}%
\special{pa 1885 1600}%
\special{pa 1890 1600}%
\special{pa 1895 1600}%
\special{pa 1900 1600}%
\special{pa 1905 1600}%
\special{pa 1910 1600}%
\special{pa 1915 1600}%
\special{pa 1920 1600}%
\special{pa 1925 1600}%
\special{pa 1930 1600}%
\special{pa 1935 1600}%
\special{pa 1940 1600}%
\special{pa 1945 1600}%
\special{pa 1950 1600}%
\special{pa 1955 1600}%
\special{pa 1960 1600}%
\special{pa 1965 1600}%
\special{pa 1970 1600}%
\special{pa 1975 1600}%
\special{pa 1980 1600}%
\special{pa 1985 1600}%
\special{pa 1990 1600}%
\special{pa 1995 1600}%
\special{pa 2000 1600}%
\special{pa 2005 1600}%
\special{pa 2010 1600}%
\special{pa 2015 1600}%
\special{pa 2020 1600}%
\special{pa 2025 1600}%
\special{pa 2030 1600}%
\special{pa 2035 1600}%
\special{pa 2040 1600}%
\special{pa 2045 1600}%
\special{pa 2050 1600}%
\special{pa 2055 1600}%
\special{pa 2060 1600}%
\special{pa 2065 1600}%
\special{pa 2070 1600}%
\special{pa 2075 1600}%
\special{pa 2080 1600}%
\special{pa 2085 1600}%
\special{pa 2090 1600}%
\special{pa 2095 1600}%
\special{pa 2100 1600}%
\special{pa 2105 1600}%
\special{pa 2110 1600}%
\special{sp}%
\special{pa 2110 1600}%
\special{pa 2115 1600}%
\special{pa 2120 1600}%
\special{pa 2125 1600}%
\special{pa 2130 1600}%
\special{pa 2135 1600}%
\special{pa 2140 1600}%
\special{pa 2145 1600}%
\special{pa 2150 1600}%
\special{pa 2155 1600}%
\special{pa 2160 1600}%
\special{pa 2165 1600}%
\special{pa 2170 1600}%
\special{pa 2175 1600}%
\special{pa 2180 1600}%
\special{pa 2185 1600}%
\special{pa 2190 1600}%
\special{pa 2195 1600}%
\special{pa 2200 1600}%
\special{pa 2205 1600}%
\special{pa 2210 1600}%
\special{pa 2215 1600}%
\special{pa 2220 1600}%
\special{pa 2225 1600}%
\special{pa 2230 1600}%
\special{pa 2235 1600}%
\special{pa 2240 1600}%
\special{pa 2245 1600}%
\special{pa 2250 1600}%
\special{pa 2255 1600}%
\special{pa 2260 1600}%
\special{pa 2265 1600}%
\special{pa 2270 1600}%
\special{pa 2275 1600}%
\special{pa 2280 1600}%
\special{pa 2285 1600}%
\special{pa 2290 1600}%
\special{pa 2295 1600}%
\special{pa 2300 1600}%
\special{pa 2305 1600}%
\special{pa 2310 1600}%
\special{pa 2315 1600}%
\special{pa 2320 1600}%
\special{pa 2325 1600}%
\special{pa 2330 1600}%
\special{pa 2335 1600}%
\special{pa 2340 1600}%
\special{pa 2345 1600}%
\special{pa 2350 1600}%
\special{pa 2355 1600}%
\special{pa 2360 1600}%
\special{pa 2365 1600}%
\special{pa 2370 1600}%
\special{pa 2375 1600}%
\special{pa 2380 1600}%
\special{pa 2385 1600}%
\special{pa 2390 1600}%
\special{pa 2395 1600}%
\special{pa 2400 1600}%
\special{pa 2405 1600}%
\special{pa 2410 1600}%
\special{pa 2415 1600}%
\special{pa 2420 1600}%
\special{pa 2425 1600}%
\special{pa 2430 1600}%
\special{pa 2435 1600}%
\special{pa 2440 1600}%
\special{pa 2445 1600}%
\special{pa 2450 1600}%
\special{pa 2455 1600}%
\special{pa 2460 1600}%
\special{pa 2465 1600}%
\special{pa 2470 1600}%
\special{pa 2475 1600}%
\special{pa 2480 1600}%
\special{pa 2485 1600}%
\special{pa 2490 1600}%
\special{pa 2495 1600}%
\special{pa 2500 1600}%
\special{pa 2505 1600}%
\special{pa 2510 1600}%
\special{pa 2515 1600}%
\special{pa 2520 1600}%
\special{pa 2525 1600}%
\special{pa 2530 1600}%
\special{pa 2535 1600}%
\special{pa 2540 1600}%
\special{pa 2545 1600}%
\special{pa 2550 1600}%
\special{pa 2555 1600}%
\special{pa 2560 1600}%
\special{pa 2565 1600}%
\special{pa 2570 1600}%
\special{pa 2575 1600}%
\special{pa 2580 1600}%
\special{pa 2585 1600}%
\special{pa 2590 1600}%
\special{pa 2595 1600}%
\special{pa 2600 1600}%
\special{sp}%
\special{pa 2600 1600}%
\special{pa 2605 1600}%
\special{pa 2610 1600}%
\special{pa 2615 1600}%
\special{pa 2620 1600}%
\special{pa 2625 1600}%
\special{pa 2630 1600}%
\special{pa 2635 1600}%
\special{pa 2640 1600}%
\special{pa 2645 1600}%
\special{pa 2650 1600}%
\special{pa 2655 1600}%
\special{pa 2660 1600}%
\special{pa 2665 1600}%
\special{pa 2670 1600}%
\special{pa 2675 1600}%
\special{pa 2680 1600}%
\special{pa 2685 1600}%
\special{pa 2690 1600}%
\special{pa 2695 1600}%
\special{pa 2700 1600}%
\special{pa 2705 1600}%
\special{pa 2710 1600}%
\special{pa 2715 1600}%
\special{pa 2720 1600}%
\special{pa 2725 1600}%
\special{pa 2730 1600}%
\special{pa 2735 1600}%
\special{pa 2740 1600}%
\special{pa 2745 1600}%
\special{pa 2750 1600}%
\special{pa 2755 1600}%
\special{pa 2760 1600}%
\special{pa 2765 1600}%
\special{pa 2770 1600}%
\special{pa 2775 1600}%
\special{pa 2780 1600}%
\special{pa 2785 1600}%
\special{pa 2790 1600}%
\special{pa 2795 1600}%
\special{pa 2800 1600}%
\special{pa 2805 1600}%
\special{pa 2810 1600}%
\special{pa 2815 1600}%
\special{pa 2820 1600}%
\special{pa 2825 1600}%
\special{pa 2830 1600}%
\special{pa 2835 1600}%
\special{pa 2840 1600}%
\special{pa 2845 1600}%
\special{pa 2850 1600}%
\special{pa 2855 1600}%
\special{pa 2860 1600}%
\special{pa 2865 1600}%
\special{pa 2870 1600}%
\special{pa 2875 1600}%
\special{pa 2880 1600}%
\special{pa 2885 1600}%
\special{pa 2890 1600}%
\special{pa 2895 1600}%
\special{pa 2900 1600}%
\special{pa 2905 1600}%
\special{pa 2910 1600}%
\special{pa 2915 1600}%
\special{pa 2920 1600}%
\special{pa 2925 1600}%
\special{pa 2930 1600}%
\special{pa 2935 1600}%
\special{pa 2940 1600}%
\special{pa 2945 1600}%
\special{pa 2950 1600}%
\special{pa 2955 1600}%
\special{pa 2960 1600}%
\special{pa 2965 1600}%
\special{pa 2970 1600}%
\special{pa 2975 1600}%
\special{pa 2980 1600}%
\special{pa 2985 1600}%
\special{pa 2990 1600}%
\special{pa 2995 1600}%
\special{pa 3000 1600}%
\special{pa 3005 1600}%
\special{pa 3010 1600}%
\special{pa 3015 1600}%
\special{pa 3020 1600}%
\special{pa 3025 1600}%
\special{pa 3030 1600}%
\special{pa 3035 1600}%
\special{pa 3040 1600}%
\special{pa 3045 1600}%
\special{pa 3050 1600}%
\special{pa 3055 1600}%
\special{pa 3060 1600}%
\special{pa 3065 1600}%
\special{pa 3070 1600}%
\special{pa 3075 1600}%
\special{pa 3080 1600}%
\special{pa 3085 1600}%
\special{pa 3090 1600}%
\special{sp}%
\special{pa 3090 1600}%
\special{pa 3095 1600}%
\special{pa 3100 1600}%
\special{pa 3105 1600}%
\special{pa 3110 1600}%
\special{pa 3115 1600}%
\special{pa 3120 1600}%
\special{pa 3125 1600}%
\special{pa 3130 1600}%
\special{pa 3135 1600}%
\special{pa 3140 1600}%
\special{pa 3145 1600}%
\special{pa 3150 1600}%
\special{pa 3155 1600}%
\special{pa 3160 1600}%
\special{pa 3165 1600}%
\special{pa 3170 1600}%
\special{pa 3175 1600}%
\special{pa 3180 1600}%
\special{pa 3185 1600}%
\special{pa 3190 1600}%
\special{pa 3195 1600}%
\special{pa 3200 1600}%
\special{pa 3205 1600}%
\special{pa 3210 1600}%
\special{pa 3215 1600}%
\special{pa 3220 1600}%
\special{pa 3225 1600}%
\special{pa 3230 1600}%
\special{pa 3235 1600}%
\special{pa 3240 1600}%
\special{pa 3245 1600}%
\special{pa 3250 1600}%
\special{pa 3255 1600}%
\special{pa 3260 1600}%
\special{pa 3265 1600}%
\special{pa 3270 1600}%
\special{pa 3275 1600}%
\special{pa 3280 1600}%
\special{pa 3285 1600}%
\special{pa 3290 1600}%
\special{pa 3295 1600}%
\special{pa 3300 1600}%
\special{pa 3305 1600}%
\special{pa 3310 1600}%
\special{pa 3315 1600}%
\special{pa 3320 1600}%
\special{pa 3325 1600}%
\special{pa 3330 1600}%
\special{pa 3335 1600}%
\special{pa 3340 1600}%
\special{pa 3345 1600}%
\special{pa 3350 1600}%
\special{pa 3355 1600}%
\special{pa 3360 1600}%
\special{pa 3365 1600}%
\special{pa 3370 1600}%
\special{pa 3375 1600}%
\special{pa 3380 1600}%
\special{pa 3385 1600}%
\special{pa 3390 1600}%
\special{pa 3395 1600}%
\special{pa 3400 1600}%
\special{pa 3405 1600}%
\special{pa 3410 1600}%
\special{pa 3415 1600}%
\special{pa 3420 1600}%
\special{pa 3425 1600}%
\special{pa 3430 1600}%
\special{pa 3435 1600}%
\special{pa 3440 1600}%
\special{pa 3445 1600}%
\special{pa 3450 1600}%
\special{pa 3455 1600}%
\special{pa 3460 1600}%
\special{pa 3465 1600}%
\special{pa 3470 1600}%
\special{pa 3475 1600}%
\special{pa 3480 1600}%
\special{pa 3485 1600}%
\special{pa 3490 1600}%
\special{pa 3495 1600}%
\special{pa 3500 1600}%
\special{pa 3505 1600}%
\special{pa 3510 1600}%
\special{pa 3515 1600}%
\special{pa 3520 1600}%
\special{pa 3525 1600}%
\special{pa 3530 1600}%
\special{pa 3535 1600}%
\special{pa 3540 1600}%
\special{pa 3545 1600}%
\special{pa 3550 1600}%
\special{pa 3555 1600}%
\special{pa 3560 1600}%
\special{pa 3565 1600}%
\special{pa 3570 1600}%
\special{pa 3575 1600}%
\special{pa 3580 1600}%
\special{sp}%
\special{pa 3580 1600}%
\special{pa 3585 1600}%
\special{pa 3590 1600}%
\special{pa 3595 1600}%
\special{pa 3600 1600}%
\special{pa 3605 1600}%
\special{pa 3610 1600}%
\special{pa 3615 1600}%
\special{sp}%
%
\special{pn 20}%
\special{pa 2600 810}%
\special{pa 1800 1600}%
\special{fp}%
\special{pa 1810 1600}%
\special{pa 2600 2400}%
\special{fp}%
\special{pa 2600 2400}%
\special{pa 3410 1600}%
\special{fp}%
\special{pa 3410 1600}%
\special{pa 2600 810}%
\special{fp}%
%
\special{pn 8}%
\special{pa 2010 1800}%
\special{pa 2790 1010}%
\special{fp}%
\special{sh 1}%
\special{pa 2790 1010}%
\special{pa 2729 1043}%
\special{pa 2753 1048}%
\special{pa 2757 1071}%
\special{pa 2790 1010}%
\special{fp}%
%
\special{pn 8}%
\special{pa 2400 2180}%
\special{pa 2780 1790}%
\special{dt 0.045}%
\special{sh 1}%
\special{pa 2780 1790}%
\special{pa 2719 1824}%
\special{pa 2743 1828}%
\special{pa 2748 1852}%
\special{pa 2780 1790}%
\special{fp}%
%
\special{pn 8}%
\special{pa 2800 1770}%
\special{pa 3180 1380}%
\special{dt 0.045}%
\special{sh 1}%
\special{pa 3180 1380}%
\special{pa 3119 1414}%
\special{pa 3143 1418}%
\special{pa 3148 1442}%
\special{pa 3180 1380}%
\special{fp}%
\put(27.5000,-7.7000){\makebox(0,0){$\pi $}}%
\put(34.9000,-14.6000){\makebox(0,0){$\pi $}}%
\put(27.6000,-24.7000){\makebox(0,0){$-\pi $}}%
\put(17.0000,-14.7000){\makebox(0,0){$-\pi $}}%
\put(25.9000,-16.1000){\makebox(0,0)[rt]{O}}%
\put(25.6000,-6.0000){\makebox(0,0)[rt]{$k_{y}$}}%
\put(36.8000,-16.6000){\makebox(0,0)[rt]{$k_{x}$}}%
%
\special{pn 8}%
\special{pa 2600 2620}%
\special{pa 2600 600}%
\special{fp}%
\special{sh 1}%
\special{pa 2600 600}%
\special{pa 2580 667}%
\special{pa 2600 653}%
\special{pa 2620 667}%
\special{pa 2600 600}%
\special{fp}%
%
\special{pn 8}%
\special{pa 1620 1600}%
\special{pa 3620 1600}%
\special{fp}%
\special{sh 1}%
\special{pa 3620 1600}%
\special{pa 3553 1580}%
\special{pa 3567 1600}%
\special{pa 3553 1620}%
\special{pa 3620 1600}%
\special{fp}%
\special{pn 8}%
\special{pa 1620 1600}%
\special{pa 1625 1600}%
\special{pa 1630 1600}%
\special{pa 1635 1600}%
\special{pa 1640 1600}%
\special{pa 1645 1600}%
\special{pa 1650 1600}%
\special{pa 1655 1600}%
\special{pa 1660 1600}%
\special{pa 1665 1600}%
\special{pa 1670 1600}%
\special{pa 1675 1600}%
\special{pa 1680 1600}%
\special{pa 1685 1600}%
\special{pa 1690 1600}%
\special{pa 1695 1600}%
\special{pa 1700 1600}%
\special{pa 1705 1600}%
\special{pa 1710 1600}%
\special{pa 1715 1600}%
\special{pa 1720 1600}%
\special{pa 1725 1600}%
\special{pa 1730 1600}%
\special{pa 1735 1600}%
\special{pa 1740 1600}%
\special{pa 1745 1600}%
\special{pa 1750 1600}%
\special{pa 1755 1600}%
\special{pa 1760 1600}%
\special{pa 1765 1600}%
\special{pa 1770 1600}%
\special{pa 1775 1600}%
\special{pa 1780 1600}%
\special{pa 1785 1600}%
\special{pa 1790 1600}%
\special{pa 1795 1600}%
\special{pa 1800 1600}%
\special{pa 1805 1600}%
\special{pa 1810 1600}%
\special{pa 1815 1600}%
\special{pa 1820 1600}%
\special{pa 1825 1600}%
\special{pa 1830 1600}%
\special{pa 1835 1600}%
\special{pa 1840 1600}%
\special{pa 1845 1600}%
\special{pa 1850 1600}%
\special{pa 1855 1600}%
\special{pa 1860 1600}%
\special{pa 1865 1600}%
\special{pa 1870 1600}%
\special{pa 1875 1600}%
\special{pa 1880 1600}%
\special{pa 1885 1600}%
\special{pa 1890 1600}%
\special{pa 1895 1600}%
\special{pa 1900 1600}%
\special{pa 1905 1600}%
\special{pa 1910 1600}%
\special{pa 1915 1600}%
\special{pa 1920 1600}%
\special{pa 1925 1600}%
\special{pa 1930 1600}%
\special{pa 1935 1600}%
\special{pa 1940 1600}%
\special{pa 1945 1600}%
\special{pa 1950 1600}%
\special{pa 1955 1600}%
\special{pa 1960 1600}%
\special{pa 1965 1600}%
\special{pa 1970 1600}%
\special{pa 1975 1600}%
\special{pa 1980 1600}%
\special{pa 1985 1600}%
\special{pa 1990 1600}%
\special{pa 1995 1600}%
\special{pa 2000 1600}%
\special{pa 2005 1600}%
\special{pa 2010 1600}%
\special{pa 2015 1600}%
\special{pa 2020 1600}%
\special{pa 2025 1600}%
\special{pa 2030 1600}%
\special{pa 2035 1600}%
\special{pa 2040 1600}%
\special{pa 2045 1600}%
\special{pa 2050 1600}%
\special{pa 2055 1600}%
\special{pa 2060 1600}%
\special{pa 2065 1600}%
\special{pa 2070 1600}%
\special{pa 2075 1600}%
\special{pa 2080 1600}%
\special{pa 2085 1600}%
\special{pa 2090 1600}%
\special{pa 2095 1600}%
\special{pa 2100 1600}%
\special{pa 2105 1600}%
\special{pa 2110 1600}%
\special{sp}%
\special{pa 2110 1600}%
\special{pa 2115 1600}%
\special{pa 2120 1600}%
\special{pa 2125 1600}%
\special{pa 2130 1600}%
\special{pa 2135 1600}%
\special{pa 2140 1600}%
\special{pa 2145 1600}%
\special{pa 2150 1600}%
\special{pa 2155 1600}%
\special{pa 2160 1600}%
\special{pa 2165 1600}%
\special{pa 2170 1600}%
\special{pa 2175 1600}%
\special{pa 2180 1600}%
\special{pa 2185 1600}%
\special{pa 2190 1600}%
\special{pa 2195 1600}%
\special{pa 2200 1600}%
\special{pa 2205 1600}%
\special{pa 2210 1600}%
\special{pa 2215 1600}%
\special{pa 2220 1600}%
\special{pa 2225 1600}%
\special{pa 2230 1600}%
\special{pa 2235 1600}%
\special{pa 2240 1600}%
\special{pa 2245 1600}%
\special{pa 2250 1600}%
\special{pa 2255 1600}%
\special{pa 2260 1600}%
\special{pa 2265 1600}%
\special{pa 2270 1600}%
\special{pa 2275 1600}%
\special{pa 2280 1600}%
\special{pa 2285 1600}%
\special{pa 2290 1600}%
\special{pa 2295 1600}%
\special{pa 2300 1600}%
\special{pa 2305 1600}%
\special{pa 2310 1600}%
\special{pa 2315 1600}%
\special{pa 2320 1600}%
\special{pa 2325 1600}%
\special{pa 2330 1600}%
\special{pa 2335 1600}%
\special{pa 2340 1600}%
\special{pa 2345 1600}%
\special{pa 2350 1600}%
\special{pa 2355 1600}%
\special{pa 2360 1600}%
\special{pa 2365 1600}%
\special{pa 2370 1600}%
\special{pa 2375 1600}%
\special{pa 2380 1600}%
\special{pa 2385 1600}%
\special{pa 2390 1600}%
\special{pa 2395 1600}%
\special{pa 2400 1600}%
\special{pa 2405 1600}%
\special{pa 2410 1600}%
\special{pa 2415 1600}%
\special{pa 2420 1600}%
\special{pa 2425 1600}%
\special{pa 2430 1600}%
\special{pa 2435 1600}%
\special{pa 2440 1600}%
\special{pa 2445 1600}%
\special{pa 2450 1600}%
\special{pa 2455 1600}%
\special{pa 2460 1600}%
\special{pa 2465 1600}%
\special{pa 2470 1600}%
\special{pa 2475 1600}%
\special{pa 2480 1600}%
\special{pa 2485 1600}%
\special{pa 2490 1600}%
\special{pa 2495 1600}%
\special{pa 2500 1600}%
\special{pa 2505 1600}%
\special{pa 2510 1600}%
\special{pa 2515 1600}%
\special{pa 2520 1600}%
\special{pa 2525 1600}%
\special{pa 2530 1600}%
\special{pa 2535 1600}%
\special{pa 2540 1600}%
\special{pa 2545 1600}%
\special{pa 2550 1600}%
\special{pa 2555 1600}%
\special{pa 2560 1600}%
\special{pa 2565 1600}%
\special{pa 2570 1600}%
\special{pa 2575 1600}%
\special{pa 2580 1600}%
\special{pa 2585 1600}%
\special{pa 2590 1600}%
\special{pa 2595 1600}%
\special{pa 2600 1600}%
\special{sp}%
\special{pa 2600 1600}%
\special{pa 2605 1600}%
\special{pa 2610 1600}%
\special{pa 2615 1600}%
\special{pa 2620 1600}%
\special{pa 2625 1600}%
\special{pa 2630 1600}%
\special{pa 2635 1600}%
\special{pa 2640 1600}%
\special{pa 2645 1600}%
\special{pa 2650 1600}%
\special{pa 2655 1600}%
\special{pa 2660 1600}%
\special{pa 2665 1600}%
\special{pa 2670 1600}%
\special{pa 2675 1600}%
\special{pa 2680 1600}%
\special{pa 2685 1600}%
\special{pa 2690 1600}%
\special{pa 2695 1600}%
\special{pa 2700 1600}%
\special{pa 2705 1600}%
\special{pa 2710 1600}%
\special{pa 2715 1600}%
\special{pa 2720 1600}%
\special{pa 2725 1600}%
\special{pa 2730 1600}%
\special{pa 2735 1600}%
\special{pa 2740 1600}%
\special{pa 2745 1600}%
\special{pa 2750 1600}%
\special{pa 2755 1600}%
\special{pa 2760 1600}%
\special{pa 2765 1600}%
\special{pa 2770 1600}%
\special{pa 2775 1600}%
\special{pa 2780 1600}%
\special{pa 2785 1600}%
\special{pa 2790 1600}%
\special{pa 2795 1600}%
\special{pa 2800 1600}%
\special{pa 2805 1600}%
\special{pa 2810 1600}%
\special{pa 2815 1600}%
\special{pa 2820 1600}%
\special{pa 2825 1600}%
\special{pa 2830 1600}%
\special{pa 2835 1600}%
\special{pa 2840 1600}%
\special{pa 2845 1600}%
\special{pa 2850 1600}%
\special{pa 2855 1600}%
\special{pa 2860 1600}%
\special{pa 2865 1600}%
\special{pa 2870 1600}%
\special{pa 2875 1600}%
\special{pa 2880 1600}%
\special{pa 2885 1600}%
\special{pa 2890 1600}%
\special{pa 2895 1600}%
\special{pa 2900 1600}%
\special{pa 2905 1600}%
\special{pa 2910 1600}%
\special{pa 2915 1600}%
\special{pa 2920 1600}%
\special{pa 2925 1600}%
\special{pa 2930 1600}%
\special{pa 2935 1600}%
\special{pa 2940 1600}%
\special{pa 2945 1600}%
\special{pa 2950 1600}%
\special{pa 2955 1600}%
\special{pa 2960 1600}%
\special{pa 2965 1600}%
\special{pa 2970 1600}%
\special{pa 2975 1600}%
\special{pa 2980 1600}%
\special{pa 2985 1600}%
\special{pa 2990 1600}%
\special{pa 2995 1600}%
\special{pa 3000 1600}%
\special{pa 3005 1600}%
\special{pa 3010 1600}%
\special{pa 3015 1600}%
\special{pa 3020 1600}%
\special{pa 3025 1600}%
\special{pa 3030 1600}%
\special{pa 3035 1600}%
\special{pa 3040 1600}%
\special{pa 3045 1600}%
\special{pa 3050 1600}%
\special{pa 3055 1600}%
\special{pa 3060 1600}%
\special{pa 3065 1600}%
\special{pa 3070 1600}%
\special{pa 3075 1600}%
\special{pa 3080 1600}%
\special{pa 3085 1600}%
\special{pa 3090 1600}%
\special{sp}%
\special{pa 3090 1600}%
\special{pa 3095 1600}%
\special{pa 3100 1600}%
\special{pa 3105 1600}%
\special{pa 3110 1600}%
\special{pa 3115 1600}%
\special{pa 3120 1600}%
\special{pa 3125 1600}%
\special{pa 3130 1600}%
\special{pa 3135 1600}%
\special{pa 3140 1600}%
\special{pa 3145 1600}%
\special{pa 3150 1600}%
\special{pa 3155 1600}%
\special{pa 3160 1600}%
\special{pa 3165 1600}%
\special{pa 3170 1600}%
\special{pa 3175 1600}%
\special{pa 3180 1600}%
\special{pa 3185 1600}%
\special{pa 3190 1600}%
\special{pa 3195 1600}%
\special{pa 3200 1600}%
\special{pa 3205 1600}%
\special{pa 3210 1600}%
\special{pa 3215 1600}%
\special{pa 3220 1600}%
\special{pa 3225 1600}%
\special{pa 3230 1600}%
\special{pa 3235 1600}%
\special{pa 3240 1600}%
\special{pa 3245 1600}%
\special{pa 3250 1600}%
\special{pa 3255 1600}%
\special{pa 3260 1600}%
\special{pa 3265 1600}%
\special{pa 3270 1600}%
\special{pa 3275 1600}%
\special{pa 3280 1600}%
\special{pa 3285 1600}%
\special{pa 3290 1600}%
\special{pa 3295 1600}%
\special{pa 3300 1600}%
\special{pa 3305 1600}%
\special{pa 3310 1600}%
\special{pa 3315 1600}%
\special{pa 3320 1600}%
\special{pa 3325 1600}%
\special{pa 3330 1600}%
\special{pa 3335 1600}%
\special{pa 3340 1600}%
\special{pa 3345 1600}%
\special{pa 3350 1600}%
\special{pa 3355 1600}%
\special{pa 3360 1600}%
\special{pa 3365 1600}%
\special{pa 3370 1600}%
\special{pa 3375 1600}%
\special{pa 3380 1600}%
\special{pa 3385 1600}%
\special{pa 3390 1600}%
\special{pa 3395 1600}%
\special{pa 3400 1600}%
\special{pa 3405 1600}%
\special{pa 3410 1600}%
\special{pa 3415 1600}%
\special{pa 3420 1600}%
\special{pa 3425 1600}%
\special{pa 3430 1600}%
\special{pa 3435 1600}%
\special{pa 3440 1600}%
\special{pa 3445 1600}%
\special{pa 3450 1600}%
\special{pa 3455 1600}%
\special{pa 3460 1600}%
\special{pa 3465 1600}%
\special{pa 3470 1600}%
\special{pa 3475 1600}%
\special{pa 3480 1600}%
\special{pa 3485 1600}%
\special{pa 3490 1600}%
\special{pa 3495 1600}%
\special{pa 3500 1600}%
\special{pa 3505 1600}%
\special{pa 3510 1600}%
\special{pa 3515 1600}%
\special{pa 3520 1600}%
\special{pa 3525 1600}%
\special{pa 3530 1600}%
\special{pa 3535 1600}%
\special{pa 3540 1600}%
\special{pa 3545 1600}%
\special{pa 3550 1600}%
\special{pa 3555 1600}%
\special{pa 3560 1600}%
\special{pa 3565 1600}%
\special{pa 3570 1600}%
\special{pa 3575 1600}%
\special{pa 3580 1600}%
\special{sp}%
\special{pa 3580 1600}%
\special{pa 3585 1600}%
\special{pa 3590 1600}%
\special{pa 3595 1600}%
\special{pa 3600 1600}%
\special{pa 3605 1600}%
\special{pa 3610 1600}%
\special{pa 3615 1600}%
\special{sp}%
%
\special{pn 20}%
\special{pa 2600 810}%
\special{pa 1800 1600}%
\special{fp}%
\special{pa 1810 1600}%
\special{pa 2600 2400}%
\special{fp}%
\special{pa 2600 2400}%
\special{pa 3410 1600}%
\special{fp}%
\special{pa 3410 1600}%
\special{pa 2600 810}%
\special{fp}%
%
\special{pn 8}%
\special{pa 2010 1800}%
\special{pa 2790 1010}%
\special{fp}%
\special{sh 1}%
\special{pa 2790 1010}%
\special{pa 2729 1043}%
\special{pa 2753 1048}%
\special{pa 2757 1071}%
\special{pa 2790 1010}%
\special{fp}%
%
\special{pn 8}%
\special{pa 2400 2180}%
\special{pa 2780 1790}%
\special{dt 0.045}%
\special{sh 1}%
\special{pa 2780 1790}%
\special{pa 2719 1824}%
\special{pa 2743 1828}%
\special{pa 2748 1852}%
\special{pa 2780 1790}%
\special{fp}%
%
\special{pn 8}%
\special{pa 2800 1770}%
\special{pa 3180 1380}%
\special{dt 0.045}%
\special{sh 1}%
\special{pa 3180 1380}%
\special{pa 3119 1414}%
\special{pa 3143 1418}%
\special{pa 3148 1442}%
\special{pa 3180 1380}%
\special{fp}%
\end{picture}%

  \caption{The situation of the Fermi-line nesting in the half-filled 
two-dimensional square-lattice system.  The thick-line square represents the 
Fermi line, and a solid line is the nesting vector connecting two different 
parts of the Fermi line. The dotted lines describe the nesting due to the 
second order perturbation, where two different parts of the Fermi line are 
connected through a second order process.} 
 \end{center}
\label{fig1}
\end{figure}
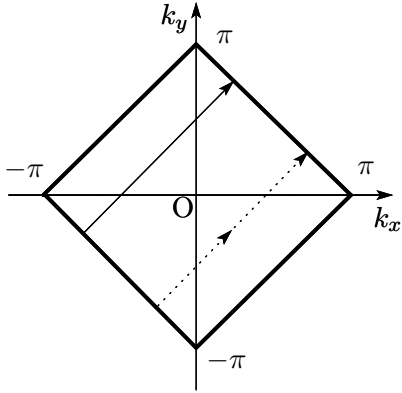

The Peierls transition~\cite{Peierls} which is caused by the competition 
between the energy loss due to lattice distortions and the gain due to the 
formation of a gap at the Fermi level in the electronic energy spectrum is 
one of peculiar properties of one-dimensional electron-lattice systems. It 
is also well known that this transition can occur even in higher dimensions 
if the structure of the Fermi surface allows strong nesting by a single 
nesting vector. A two-dimensional square lattice system with a half-filled 
electronic band described by a tight-binding model is a typical example 
where the Fermi surface (more properly Fermi line) can be completely nested 
by the wave vector $\vv{Q}=(\pi/a,\pi/a)$ with $a$ the lattice constant. 
Early studies on the issue of two-dimensional Peierls transition in the 
half-filled systems suggested that the lowest energy phase would be the 
state with lattice distortions whose Fourier components involved only 
$\vv{Q}$-modes.~\cite{Tang} The relation between the electron-electron 
interactions and the two-dimensional Peierls instability has been discussed 
rather intensively in connection to the mechanism of high $T_{\rm c}$ 
superconductivity.~\cite{Tang,Machida87,Scal89,Mazumdar,Tang89} Nevertheless 
the investigation of the true ground state of the pure electron-lattice 
system without electron interactions was not clearly understood until 
recently. In recent works by Hamano and one of the present 
authors,~\cite{Ono00,Hamano01} it has been shown that the lowest energy 
state of a two-dimensional electron-lattice system described by a 
two-dimensional version of Su, Schrieffer and Heeger's (SSH) model 
Hamiltonian~\cite{SSH} involves not only the distortions with 
$\vv{Q}$-vector but also those with various wave vectors parallel to 
$\vv{Q}$, the latter contributing to the formation of the electronic energy 
gap through a second order process (see Fig.~\ref{fig1}), and that there are 
an infinite number of non-equivalent lattice distortion patterns which 
indicates an infinite degeneracy of the lowest energy states. It has also 
been argued~\cite{Hamanox} that those behaviors are maintained at finite 
temperatures and that the amplitudes of lattice distortions with different 
wave vectors, all being parallel to $\vv{Q}$, vanish all together 
at a critical temperature. 

It is also known~\cite{Schulz78} that the Peierls transition in 
one-dimensional systems is accompanied by the softening of the 2$k_{\rm F}$ 
phonon mode at the transition temperature, where $k_{\rm F}$ represents the 
Fermi wave number and 2$k_{\rm F}$ corresponds to $\vv{Q}$ in the 
two-dimensional square lattice systems with a half-filled electronic band. 
As has been clarified by previous papers, phonon modes relevant to the 
Peierls transition in the two-dimensional square lattice system described 
by the SSH model involve not only the $\vv{Q}$ modes but also many other 
modes with wave vectors parallel to $\vv{Q}$. Furthermore, as discussed in 
refs.~\citen{Ono00} and \citen{Hamano01}, the lowest energy states are not 
unique. In this sense it will be worthwhile to investigate the phonon 
dispersion at finite temperatures in this two-dimensional system. Even at 
temperatures higher than the transition temperature, we may expect a phonon 
softening phenomenon which is much different from that in one-dimension. In 
the present paper, we consider the phonon dispersion of the two-dimensional 
square lattice system with a half-filled electronic band only in the 
temperature region higher than the critical one, and study which types of 
phonon modes are softened when the temperature approaches the critical one 
from above. The dispersion at lower temperatures cannot be treated in a 
simple way because of the degeneracy of the lowest energy states, and 
therefore it will be discussed in a forthcoming separate paper. 

Since the formation of the energy gap at the Fermi level leads a metallic 
system into an insulator, it is quite important to understand the 
mechanism of the Peierls transition. The present work will shed light on 
the understanding of the Peierls instability in two-dimensional 
electron-phonon systems. 

The paper is organized as follows; in the following section the model 
for and the method of the calculation are described. In Sec. 3, the 
results of the calculations are shown. The last section is devoted to 
summary and discussion.

\section{Model and Formulation} \label{sec2}

In the following calculations we use a two-dimensional version of the SSH 
model~\cite{SSH} which was originally introduced in the studies of 
one-dimensional electron-lattice systems and where the electron-lattice 
coupling is introduced through the dependence of the electronic 
nearest-neighbor transfer integral on the hopping distance. Since we are 
treating a two-dimensional square lattice, the explicit form of the model 
Hamiltonian is written as follows,
\begin{eqnarray}
H & = & -\sum _{i,j,s} \Biggl\{  \big[ t_{0}-\alpha \big( 
u_{x}(i+1,j)-u_{x}(i,j) \big) \big] \nonumber \\
& & \times \left(c_{i+1,j,s}^{\dagger }c_{i,j,s}+c_{i,j,s}^{\dagger }c_{i+1,j,s}
\right) \nonumber \\
& & + \big[ t_{0}-\alpha \big( u_{y}(i,j+1)-u_{y}(i,j)\big) \big] 
\nonumber \\
& & \times \left(c_{i,j+1,s}^{\dagger }c_{i,j,s}+c_{i,j,s}^{\dagger }c_{i,j+1,s}
\right) \Biggr\} \nonumber \\
& & +\frac{K}{2}\sum _{i,j}  \bigg[ \big( u_{x}(i+1,j)-u_{x}(i,j)\big)^{2} 
\nonumber \\
& & + \big( u_{y}(i,j+1)-u_{y}(i,j)\big) ^{2}\bigg] \nonumber \\
& & +\frac{M}{2}\sum _{i,j}  \bigg[ \big( \dot{u}_{x}(i,j)\big)^2+
\big( \dot{u}_y(i,j)\big)^2 
\bigg], \label{eq2.1}
\end{eqnarray}
where the field operators $c_{i,j,s}$ and $c_{i,j,s}^{\dagger }$ annihilate 
and create an electron with spin $s$ at the site $(i,j)$, respectively,  and 
$t_{0}$ is the transfer integral for the equidistant lattice, $\alpha$ the 
electron-lattice coupling constant, $\vv{u} (i,j)$ the lattice 
displacement vector whose $x$- and $y$-components are denoted as $u_x(i,j)$ 
and $u_y(i,j)$, respectively, $K$ the force constant describing ionic 
coupling strength in the lattice system, $M$ the mass of an ion unit at a 
site. Through out this paper, we assume the periodic boundary conditions 
(PBC) for both directions.

The equations of motion for the lattice system can be derived on the 
assumption that the ionic mass $M$ is much larger than the effective mass 
of an electron and by introducing the Born-Oppenheimer 
approximation;~\cite{Born27} in the case of the $x$-components of the 
displacement vectors, they are explicitly written as follows,
\begin{eqnarray}
M\ddot{u} _{x}(i,j)& = & \alpha \sum_s \big( 
\langle c^{\dagger}_{i+1,j,s}c_{i,j,s}
+c^{\dagger}_{i,j,s}c_{i+1,j,s} \rangle \nonumber \\
& & -\langle c_{i,j,s}^{\dagger }c_{i-1,j,s}+
c_{i-1,j,s}^{\dagger}c_{i,j,s}\rangle \big) \nonumber \\
& & +K \big( u_{x}(i+1,j)-2u_x(i,j)+u_{x}(i-1,j) \big), \label{eq2.2}
\end{eqnarray}
and similar expressions for the $y$ components can be written down, which 
will be omitted here because they can be easily speculated from the above 
ones. 
Assuming the electronic system is in the equilibrium state for the given 
lattice configuration at any moment, the above equations of motion can be 
rewritten as follows,
\begin{eqnarray}
M\ddot{u}_{x}(i,j) & = & 2\alpha \sum_{\nu} f(\varepsilon_{\nu }) 
\biggl( \phi_{\nu }(i+1,j)\phi_{\nu }(i,j) \nonumber \\
& & -\phi_{\nu }(i,j)\phi_{\nu}(i-1,j) \biggr) \nonumber \\
& & +K \big( u_{x}(i+1,j)-2u_x(i,j)+u_{x}(i-1,j) \big), \label{eq2.3}
\end{eqnarray}
where $f(\varepsilon_{\nu})$ represents the Fermi distribution function for 
the eigenenergy $\varepsilon_\nu$ corresponding to the single particle 
eigenstate $\phi_\nu$, which is obtained by solving the eigenvalue problem 
for a given configuration of the lattice displacements \{ $\vv{u}(i,j)$\}. 
Because of the electron-hole symmetry of the present system, the electronic 
chemical potential is fixed to be zero at any temperature. Again the equations 
for the $y$-component are omitted. 

The stable static lattice displacements \{ $\vv{u}^0(i,j)$ \} are obtained 
by solving a set of self-consistent equations which are nothing but those 
given by setting the right-hand side of eq.~(\ref{eq2.3}) to be zero. The 
linear modes around this static solution, i.e. the phonon modes, can be 
calculated by linearizing the equations of motion for $u_x(i,j)$ and 
$u_y(i,j)$ with respect to the deviations from the static configurations. 
In deriving equations for the linear modes, we should take into account the 
effects of those deviations on the electronic eigenfunctions:~\cite{Terai86} 
\begin{eqnarray}
\vv{u}(i,j,t) & = & \vv{u}^{0}(i,j)+\delta u(i,j,t) \label{eq2.4} \\
\phi_{\nu }(i,j;t)& = & \phi_{\nu }^{0}(i,j)+\delta \phi_{\nu }(i,j;t), 
\label{eq2.5}
\end{eqnarray}
where we have introduced the time variable $t$ explicitly and the electronic 
eigenfunctions corresponding to the static displacements are denoted as 
$\phi_\nu^0(i,j)$. The time dependence of the electronic wave functions is 
implicit one through the {\it slow} time dependence of the lattice 
displacements. The deviations of the wave functions $\delta \phi_{\nu }(i,j;t)$ 
are calculated within the first order perturbation theory in the form, 
\begin{equation}
\delta \phi_{\nu }(i,j;t) = \sum_{\mu (\ne \nu)} 
\frac{\phi_\mu^0(i,j)}{\varepsilon_{\nu}-\varepsilon_{\mu}}V_{\nu \mu}, 
\label{eq2.6}
\end{equation}
with
\begin{eqnarray}
V_{\nu \mu} & = & 2\alpha \sum_{n,m}
\Bigg[ \biggl( \delta u_{x}(n+1,m,t)-\delta u_{x}(n,m,t)\biggr) \nonumber \\
& & \times \biggl( \phi_{\mu}^0(n+1,m)\phi_{\nu}^0(n,m)\biggr) \nonumber \\
& & +\biggl( \delta u_y(n,m+1,t)-\delta u_y(n,m,t)\biggr) \nonumber \\
& & \times \biggl( \phi_{\mu}^0(n,m+1)\phi_{\nu}^0(n,m) \biggr) \Bigg] .
\label{eq2.7}
\end{eqnarray}

In order to determine the normal modes, we postulate that the time dependence 
of the deviations of the lattice displacements is given by 
\begin{equation}
\delta \vv{u}(i,j,t)=(\delta u_x(i,j,t),\delta u_y(i,j,t))=
\delta \vv{u}(i,j;\omega){\rm e}^{{\rm i}\omega t}. \label{eq2.8}
\end{equation}
Then the equation for $\delta \vv{u}(i,j;\omega)$ can be expressed in the 
following form,
\begin{equation}
M\omega^2 \delta \vv{u}(i,j;\omega) = \sum_{m,n} {\cal W}(i,j;m,n)\delta 
\vv{u}(m,n;\omega), \label{eq2.9}
\end{equation}
where the 2$\times$2 matrices ${\cal W}(i,j;m,n)$ are defined as follows, 
\begin{eqnarray}
{\cal W}_{a,b}(\vv{r};\vv{R}) & = & 2\alpha ^{2}\sum_{\nu}\sum_{\mu (\ne \nu)}
\frac{f(\varepsilon_{\nu })}{\varepsilon_{\nu}-\varepsilon_{\mu}} \nonumber \\
& & \times \biggl[ \phi_{\mu}^0(\vv{r})\biggl( \phi_{\nu}^0(\vv{r}+\vv{e}_a)
-\phi_{\nu}^0(\vv{r}-\vv{e}_a)\biggr) \nonumber \\
& &+\phi_{\nu}^0(\vv{r})\biggl( \phi_{\mu}^0(\vv{r}+\vv{e}_a)
-\phi_{\mu}^0(\vv{r}-\vv{e}_a)\biggr) \biggr] \nonumber \\
& & \times \Biggl[ \phi_{\mu}^0(\vv{R})
\biggl(  \phi_{\nu}^0(\vv{R}+\vv{e}_b)-\phi_{\nu}^0(\vv{R}-\vv{e}_b)\biggr)  
\nonumber \\
& & +\phi_{\nu}^0(\vv{R})\biggl(  \phi_{\mu}^0(\vv{R}+\vv{e}_b)
-\phi_{\mu}^0(\vv{R}-\vv{e}_b)\biggr) \Biggr] \nonumber \\
& & +K\delta_{a,b}\big( -\delta_{\vs{r}+\vs{e}_a,\vs{R}}
+2\delta_{\vs{r},\vs{R}}-\delta_{\vs{r}-\vs{e}_a,\vs{R}}\big) . 
\label{eq2.10}
\end{eqnarray}
Here $a$ and $b$ stand for $x$ or $y$, and for simplicity we expressed the 
lattice sites as $\vv{r}[=(i,j)]$ and $\vv{R}[=(m,n)]$ and introduced unit 
vectors $\vv{e}_x\equiv (1,0)$ and $\vv{e}_y\equiv (0,1)$. The eigenfrequency 
$\omega$ and the corresponding eigenfunction $\delta \vv{u}(\vv{r};\omega)$ 
for the normal modes are determined by solving the eigenvalue problem 
eq.~(\ref{eq2.9}). 

The above-mentioned formulation can be used at any temperature irrespective  
of the presence or absence of finite static lattice distortions. In the 
present paper we consider only the high temperature region where there is no 
static lattice distortion. In this situation, the normal modes equations 
can be reduced into a much simpler form. If there is no distortion, the 
electronic eigenenergies and eigenfunctions are given by those of the plane 
waves;
\begin{eqnarray}
\varepsilon_{\vs{k}} & = & -2t_0 [\cos(k_x)+\cos(k_y)] , \label{eq2.11} \\
\phi_{\vs{k}}^0(\vv{r}) & = & \frac{1}{L}{\rm e}^{{\rm i}\vs{k}\cdot \vv{r}}, 
\label{eq2.12}
\end{eqnarray}
where $\vv{k}=(k_x,k_y)$ is the wave vector characterizing the electronic 
eigenstate, and the system size is assumed to be $L\times L$. We use a length 
unit where the lattice constant is equal to unity. It can be easily confirmed 
that the phonon normal modes are also expressed in the plane wave form in the 
present situation, i.e. $\delta \vv{u}(\vv{r},\omega) = 
\vv{g}(\vv{q},\omega){\rm e}^{{\rm i}\vs{q}\cdot \vv{r}}$. 
Thus the linear modes equations are rewritten in the following form,
\begin{equation}
\omega^2 \vv{g}(\vv{q},\omega) = {\cal U}(\vv{q})\vv{g}(\vv{q},\omega), 
\label{eq2.13}
\end{equation}
where the elements of the 2$\times$2 matrix ${\cal U}(\vv{q})$ are given by 
\begin{eqnarray}
{\cal U}_{a,b}(\vv{q})& = & \frac{4\alpha^2}{ML^2} \sum_{\vs{k}} 
\frac{f(\varepsilon_{\vs{k}})-f(\varepsilon_{\vs{k}+\vv{q}})}
{\varepsilon_{\vs{k} }-\varepsilon_{\vs{k}+\vs{q}}} \nonumber \\
& & \times \biggl( \sin(k_a+q_a)- \sin k_a \biggr) 
\biggl( \sin (k_b+q_b)-\sin k_b \biggr) \nonumber \\ 
& & +\frac{K}{M} \bigl( 1-\cos q_a \bigr) \delta_{a,b}. \label{eq2.14}
\end{eqnarray}
Here again $a$ and $b$ stand for $x$ or $y$. It is easily confirmed that 
${\cal U}(\vv{q})$ is a symmetric matrix. The eigenfrequency $\omega$ does 
not appear in the matrix ${\cal U}(\vv{q})$ because of the Born-Oppenheimer 
approximation. It is straightforward to 
calculate the eigenvalues and eigenvectors from eq.~(\ref{eq2.13}). The 
results are discussed in the following section. It should be noted that 
the calculation of the phonon dispersion in the low temperature region, 
where distortions involving various wave numbers parallel to the nesting 
vector $\vv{Q}$ are appearing, is not so simple as in the present situation 
without any static distortion. 

\section{Results of Calculations}

As stated in the previous section, the frequencies of the phonon modes 
in the temperature region higher than the Peierls transition temperature 
$T_{\rm c}$ are obtained from the eigenvalue problem for a 2$\times$2 
matrix ${\cal U}$, the elements of which depend only on the wave vector 
$\vv{q}$ and the temperature included in the electronic distribution 
function (Fermi distribution). The result is written as
\begin{equation}
\omega^2 = \frac{1}{2} \left\{ {\cal U}_{x,x}+{\cal U}_{y,y} 
\pm \sqrt{({\cal U}_{x,x}-{\cal U}_{y,y})^2+4{\cal U}_{x,y}^2} \right\} 
\label{eq3.1}
\end{equation}

When $q_x=q_y=0$, all the elements of the matrix ${\cal U}$ vanishes 
irrespective of the temperature and consequently we have two degenerate 
uniform modes with zero frequency; they correspond to the long-wave-length 
limit of two acoustic modes. For general values of $q_x$ and $q_y$, the 
matrix elements are calculated only numerically. 

\begin{figure}[htb]
 \begin{center}
    \resizebox{80mm}{!}{\includegraphics{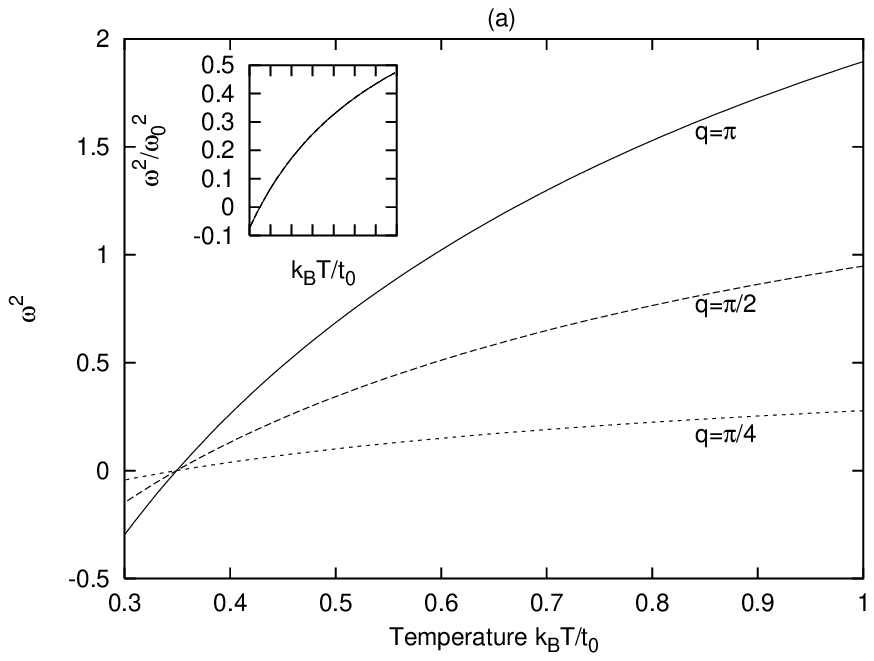}}
    \resizebox{80mm}{!}{\includegraphics{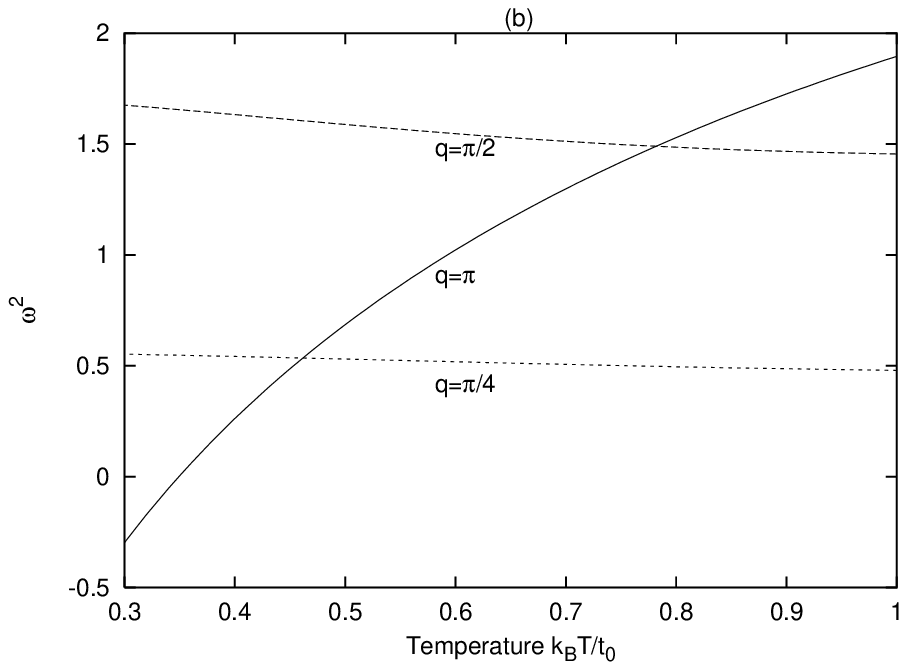}}
  \caption{The temperature dependence of the eigenvalue $\omega ^{2}$ 
for transverse (a) and longitudinal (b) modes with fixed values of the wave 
number $q~(=q_x=q_y)$; the values of $q$ are given in the graph. 
In the inset of (a) all eigenvalues for the transverse modes with $q_{x}=q_{y}=q$ 
(including $q=\pi $ ) scaled by corresponding values for free phonons $\omega _{0}^2(\vv{q} )$ 
are plotted; they fall into a single curve.
The temperature ($k_{\rm B}T$ with $k_{\rm B}$ the Boltzmann constant) is scaled 
by $t_0$ and $\omega^2$ by $K/M$. The dimensionless coupling constant is 
$\lambda =0.32$ and the system size $L$ is chosen to be 64. }
\end{center}
\label{fig2x}
\end{figure}

\begin{figure}[htb]
\begin{center}
  \resizebox{115mm}{!}{\includegraphics{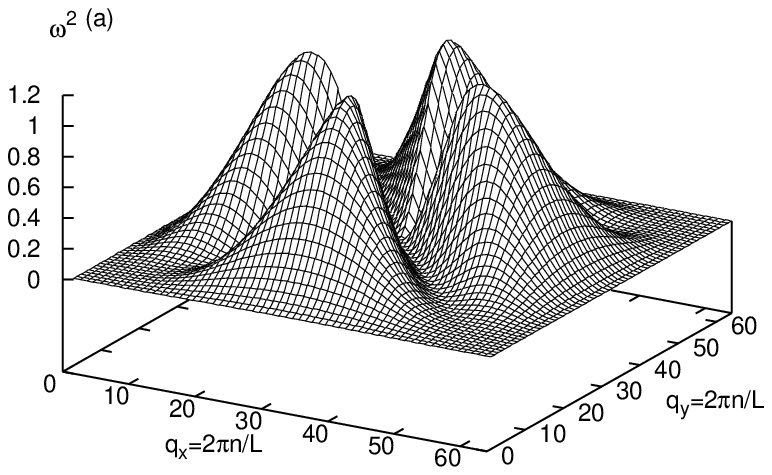}} 
  \resizebox{115mm}{!}{\includegraphics{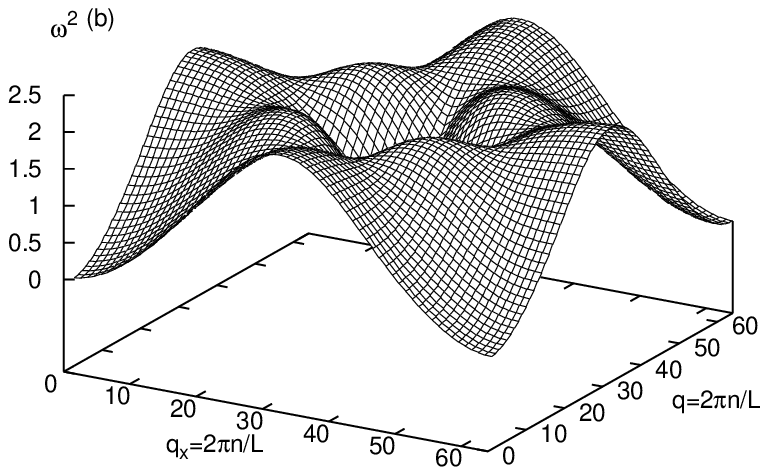}}
\caption{Phonon dispersions at $T_{c}$. (a) the dispersion related to 
``transverse'' modes and (b) that related to ``longitudinal'' modes. 
The eigenvalue $\omega^2$ is scaled by $K/M$. The dimensionless coupling 
constant $\lambda$ is 0.32, the system size $L$ being 64. }
\end{center}
\label{fig3}
\end{figure}

As has been shown in the previous works,~\cite{Ono00,Hamano01} the phonon 
modes relevant to the Peierls transition are those with wave vectors parallel 
to $\vv{Q}$ including $\vv{Q}$ itself. Therefore, let us first discuss the 
temperature dependence of several phonon frequencies for the case with 
$q_x=q_y=q$. In this case, the phonon modes are longitudinal 
($\vv{g}(\vv{q}) \parallel \vv{q}$, i.e. $g_x=g_y$) or transverse 
($\vv{g}(\vv{q}) \perp \vv{q}$, i.e. $g_x=-g_y$).~\cite{AM} 
In Fig.~\ref{fig2x} the eigenvalue $\omega^2$ is plotted 
as a function of the temperature $T$ for the transverse (a) and longitudinal 
(b) modes with $q=\pi$ (solid curve), $\pi/2$ (dashed curve) and $\pi/4$ 
(dotted curve). The product of the temperature $T$ and the Boltzmann constant 
is scaled by $t_0$ and $\omega^2$ by $K/M$. The dimensionless coupling 
constant $\lambda \equiv \alpha^2/t_0K$ is assumed to be 0.32 throughout the 
paper. In evaluating the $\vv{k}$-sums included in the definition of the 
matrix elements ${\cal U}_{a,b}$, we have set the system size $L$ to be 64; 
this choice of the system size is to get a clear presentation of the wave 
vector dependence of the dispersion (see Fig.~\ref{fig3}). It has been 
confirmed that the results shown in Fig.~\ref{fig2x} do not change even if 
larger values of $L$ are used. 

In Fig.~\ref{fig2x}(a), there are three examples of temperature dependence of the 
eigenfrequencies for the transverse modes with wavevectors $q_{x}=q_{y}=q$ 
are plotted, it should be noted that all the eigenvalues cross 0 at 
the same temperature ($k_{\rm B}T_{\rm c} \simeq 0.35$). 
We have numerically confirmed that the eigenvalues $\omega ^2(\vv{q} ,T)$ for 
the transverse modes with $q_{x}=q_{y}=q$ scaled by $\omega _{0}^2 $ $[=(K/M)(1-\cos \vv{q})]$ 
the square frequency of the corresponding free phonon depend on the temperature $T$ but not on 
$q$ as shown in the inset of Fig.2 (a).
It will be clear that all the frequencies 
of the transverse modes with wavevectors $\vv{Q}$ and the  parallel to it vanish at the same 
temperature.
This is consistent with the structure of the lower temperature Peierls 
phase.~\cite{Ono00,Hamano01,Hamanox} The critical temperature $T_{\rm c}$ is 
the same with that estimated from the vanishing of the order parameters in 
the low temperature regime.~\cite{Hamanox} In the case of the longitudinal 
modes, only the mode with the nesting vector (i.e. $q=\pi$) vanishes at the 
critical temperature as seen in Fig.~\ref{fig2x}(b).
When the temperature is 
reduced below the critical temperature, the values of $\omega^2$ for all the 
modes in Fig.~\ref{fig2x}(a) and that for the nesting vector mode in 
Fig.~\ref{fig2x}(b) become negative, indicating these modes are unstable in 
the lower temperature regime and suggesting a phase transition. 

In order to see the softening of the phonon modes with wave vectors parallel 
to the nesting vector at the critical temperature, we have calculated the 
wave vector dependence of the phonon dispersion at $T_{\rm c}$. The results 
are summarized in Fig.~\ref{fig3}, where in (a) the dispersion of the lower 
branch [see eq.~(\ref{eq3.1})] which corresponds to the transverse modes for 
some special directions of the wave vector is shown and in (b) the upper 
branch connected to the longitudinal modes for some special directions. 

For the sake of simplicity we call the lower branch [Fig.~\ref{fig3}(a)] 
the {\it transverse} modes and the upper branch [Fig.~\ref{fig3}(b)] the 
{\it longitudinal} modes. In the case of the {\it transverse} modes, all 
those with wave vectors parallel to the nesting vector $\vv{Q}$, including 
the $\vv{Q}$-mode itself, are found to have zero frequency at the critical 
temperature, while in the case of the {\it longitudinal} modes, only the 
frequency of the $\vv{Q}$-mode vanishes at the critical temperature.

\section{Summary and Discussion} \label{sec4}

In the present work, the temperature dependence of the phonon dispersion 
in the high temperature regime (i.e. at temperatures higher than the 
Peierls transition temperature) 
has been studied for a two-dimensional electron-lattice system with a 
half-filled electronic band, which is described by the SSH-type model 
extended to the case of a two-dimensional square lattice. In this system, 
it was shown in the previous papers~\cite{Ono00,Hamano01,Hamanox} that the 
lowest energy state in the low temperature regime is the Peierls state with 
multi-modes distortions where the Fourier components of the distortions 
involve not only the nesting vector modes but also many other modes having 
wave vectors parallel to the nesting vector $\vv{Q}=(\pi,\pi)$. In accordance 
with this low temperature behavior of the system, we have found the softening 
of phonon modes with wave vectors $\vv{Q}$ and with those parallel to it as 
the temperature is lowered from the high temperature region towards the critical 
temperature. As for the polarization of the softened modes, both of the 
longitudinal and transverse modes are softened in the case of the 
$\vv{Q}$-vector, while only the transverse mode has zero frequency in the case 
of wave vectors parallel to but not equal to $\vv{Q}$ at the critical 
temperature. 

Although the zero temperature ground state of the system has an infinite number 
of degeneracy,~\cite{Hamano01} depending on different amplitude distributions 
of the Fourier components of the {\it static} Peierls distortions, we have 
confirmed that the polarization of distortion components corresponding to 
non-$\vv{Q}$-modes (i.e. those with wave vectors parallel but not equal to 
$\vv{Q}$) is transverse and that the polarization of the $\vv{Q}$-mode is 
longitudinal, in all cases. This fact is maintained even at finite 
temperatures (below the critical temperature). In spite of the softening of 
the transverse $\vv{Q}$-mode, there is no transverse component of the 
$\vv{Q}$-mode in the ground state Peierls distortions. On the other hand, the 
asymmetrically dimerized Peierls state, which was suggested by Tang and 
Hirsch~\cite{Tang} but has a higher energy than the multi-modes Peierls 
state,~\cite{Ono00} involves the longitudinal and transverse components of 
the $\vv{Q}$-mode with an equal amplitude in the Fourier expansion of the 
distortions. 

Similarly, although all the transverse modes with wave vectors parallel to 
$\vv{Q}$ are softened, not all the components appear in the Fourier expansion 
of the multi-modes Peierls distortions. The simplest pattern in the degenerate 
ground states will be the one consisting of the longitudinal $\vv{Q}$-mode and 
the transverse $\vv{Q}/2$-mode.~\cite{Hamano01} These facts indicate that, 
when many phonon modes are softened, it is not necessary for all the modes to 
show condensation in order to realize the lowest energy Peierls state. At the 
moment, there is no guiding principle to determine which of the infinitely 
degenerate low temperature states is chosen when the temperature is lowered 
from the higher region across the critical temperature. 

It will also be worthwhile to study the phonon dispersion in the Peierls 
distorted phase, which is to be treated in the future work. Furthermore the 
effect of electron-electron interactions should be considered in order to 
discuss the Peierls transition in real systems, which is left for future 
studies.

\section*{Acknowledgements}

The authors are grateful to T. Hamano and H. Watanabe for useful comments 
and discussion. This work was partially supported by a Grant in Aid for 
Scientific Research (No. 14540365) from the Ministry of Education, Culture, 
Sports, Science and Technology.

\def\jpsj{J. Phys. Soc. Jpn. }
\def\ptp{Prog. Theor. Phys. }
\def\prl{Phys. Rev. Lett. }
\def\pr{Phys. Rev. }

\end{document}